\documentstyle[epsfig]{aprim}

\newif\ifAMStwofonts

\ifoldfss
  \ifCUPmtlplainloaded \else
    \NewTextAlphabet{textbfit} {cmbxti10} {}
    \NewTextAlphabet{textbfss} {cmssbx10} {}
    \NewMathAlphabet{mathbfit} {cmbxti10} {} 
    \NewMathAlphabet{mathbfss} {cmssbx10} {} 
  \fi
  \ifAMStwofonts
    \ifCUPmtlplainloaded \else
      \NewSymbolFont{upmath} {eurm10}
      \NewSymbolFont{AMSa} {msam10}
      \NewMathSymbol{\upi}     {0}{upmath}{19}
      \NewMathSymbol{\umu}     {0}{upmath}{16}
      \NewMathSymbol{\upartial}{0}{upmath}{40}
      \NewMathSymbol{\leqslant}{3}{AMSa}{36}
      \NewMathSymbol{\geqslant}{3}{AMSa}{3E}

    \fi
  \fi
\fi 

\ifnfssone
  \newmathalphabet{\mathit}
  \addtoversion{normal}{\mathit}{cmr}{m}{it}
  \addtoversion{bold}{\mathit}{cmr}{bx}{it}
  \newmathalphabet{\mathbfit} 
  \addtoversion{normal}{\mathbfit}{cmr}{bx}{it}
  \addtoversion{bold}{\mathbfit}{cmr}{bx}{it}
  \newmathalphabet{\mathbfss} 
  \addtoversion{normal}{\mathbfss}{cmss}{bx}{n}
  \addtoversion{bold}{\mathbfss}{cmss}{bx}{n}
  \ifAMStwofonts
    \ifCUPmtlplainloaded \else
      \UseAMStwoboldmath
      \makeatletter
      \new@mathgroup\upmath@group
      \define@mathgroup\mv@normal\upmath@group{eur}{m}{n}
      \define@mathgroup\mv@bold\upmath@group{eur}{b}{n}
      \edef\UPM{\hexnumber\upmath@group}
      \new@mathgroup\amsa@group
      \define@mathgroup\mv@normal\amsa@group{msa}{m}{n}
      \define@mathgroup\mv@bold\amsa@group{msa}{m}{n}
      \edef\AMSa{\hexnumber\amsa@group}
      \makeatother
      \mathchardef\upi="0\UPM19
      \mathchardef\umu="0\UPM16
      \mathchardef\upartial="0\UPM40
      \mathchardef\leqslant="3\AMSa36
      \mathchardef\geqslant="3\AMSa3E
    \fi
  \fi
\fi 

\ifnfsstwo
  \DeclareMathAlphabet{\mathbfit}{OT1}{cmr}{bx}{it}
  \SetMathAlphabet\mathbfit{bold}{OT1}{cmr}{bx}{it}
  \DeclareMathAlphabet{\mathbfss}{OT1}{cmss}{bx}{n}
  \SetMathAlphabet\mathbfss{bold}{OT1}{cmss}{bx}{n}
  \ifAMStwofonts
    \ifCUPmtlplainloaded \else
      \DeclareSymbolFont{UPM}{U}{eur}{m}{n}
      \SetSymbolFont{UPM}{bold}{U}{eur}{b}{n}
      \DeclareSymbolFont{AMSa}{U}{msa}{m}{n}
      \DeclareMathSymbol{\upi}{0}{UPM}{"19}
      \DeclareMathSymbol{\umu}{0}{UPM}{"16}
      \DeclareMathSymbol{\upartial}{0}{UPM}{"40}
      \DeclareMathSymbol{\leqslant}{3}{AMSa}{"36}
      \DeclareMathSymbol{\geqslant}{3}{AMSa}{"3E}
    \fi
  \fi
\fi 

\ifCUPmtlplainloaded \else
  \ifAMStwofonts \else 
    \def\upi{\pi}
    \def\umu{\mu}
    \def\upartial{\partial}
  \fi
\fi

\title[Manuscript Template]{The MOA 1.8-metre
alt-az wide-field survey telescope and the MOA project}

\author[Hearnshaw et al.]
       {J.B. Hearnshaw$^1$, F. Abe$^2$, I.A. Bond$^3$, A.C. Gilmore$^1$,
Y. Itow$^2$, K. Kamiya$^2$, \newauthor K. Masuda$^2$, Y. Matsubara$^2$, Y. Muraki$^2$, C.
Okada$^2$, N.J. Rattenbury$^4$, T. Sako$^2$, \newauthor M. Sasaki$^2$, D.J. Sullivan$^5$,
P.C.M. Yock$^6$
\\
        $^1$Department of Physics and Astronomy, University of Canterbury, New Zealand\\
        $^2$Solar-Terrestrial Environment Laboratory, Nagoya University, Nagoya, Japan\\
        $^3$Institute of Information and Mathematical Sciences, Massey University at Albany, New Zealand\\
        $^4$Jodrell Bank Observatory, Macclesfield, U.K.\\
        $^5$School of Chemical and Physical Sciences, Victoria University of Wellington, New Zealand\\
        $^6$Faculty of Science, University of Auckland, New Zealand}
\date{}

\pagerange{\pageref{firstpage}--\pageref{lastpage}}
\pubyear{2005}

\begin{document}

\maketitle

\label{firstpage}

\begin{abstract}
A new 1.8-m wide-field alt-az survey telescope was installed at Mt John University
Observatory in New Zealand in October 2004. The telescope will be dedicated to the MOA
(Microlensing Observations in Astrophysics) project. The instrument is equipped with a
large 10-chip mosaic CCD camera with 80 Mpixels covering about 2 square degrees of sky. It
is mounted at the f/3 prime focus. The telescope will be used for finding and following
microlensing events in the galactic bulge and elsewhere, with an emphasis on the analysis
of microlensing light curves for the detection of extrasolar planets. The MOA project is a
Japan-New Zealand collaboration, with the participation of Nagoya University and four
universities in New Zealand.
\end{abstract}


\section{Introduction}

The MOA (Microlensing Observations in Astrophysics) project was
established in 1995 as a joint project between Nagoya University,
Japan, and the universities of Auckland, Canterbury and Victoria and
Carter Observatory in New Zealand. Massey University at Albany, NZ, is
now also a member. About 20 scientists are involved in several
institutions in Japan and NZ (but mainly Nagoya, Auckland, Canterbury,
Victoria and Massey universities). All observations for MOA have so far
been made with large CCD cameras mounted on the Boller and Chivens
61-cm  f/6.25 Cassegrain telescope at Mt John University Observatory of
the University of Canterbury. The coordinates are latitude
$-43^{\circ}\,59.2^{\prime}$S, longitude $170^{\circ}\,27.9^{\prime}$E,
altitude 1025 m.

Current MOA research includes CCD observations of crowded star fields
in the galactic bulge and in the Magellanic Clouds to detect
gravitational microlensing events. Difference imaging software is used
as this is ideal for finding stars that have changed in brightness in
crowded star fields. MOA issues microlensing alerts to the
international community (see www.physics.auckland.ac.nz/moa/) and also
makes follow-up observations of high magnification and other
microlensing events of interest. Although the primary goal of MOA is
the detection and characterization of extrasolar planets in
microlensing events, secondary goals are the analysis of finite source
microlensing events, the detection of planets by photometric transits,
the search for dark matter, the analysis of variable stars and the
detection of optical afterglows from gamma-ray bursts.

In June 2002, one of us (Muraki) was successful in obtaining a grant
for a new larger telescope for MOA observations at Mt John. The grant
provides funds over 5 years (2003-2007) for the telescope, large CCD
camera, dome and initial observations. The telescope was constructed by
the Nishimura Company in Kyoto, Japan in 2003-2004. Design requirements
were a large field of view $\sim 1.5^{\circ}$, a fast focal ratio (f/3
was chosen) and good imaging over approximately 380 nm to 1 $\mu$m.

The optical design was undertaken by A.\ Rakich (IRL, Lower Hutt, N.Z.)
and incorporates an f/2.75 primary mirror, giving a focal ratio of
f/2.91 at the prime focus after passage through the four-lens
aberration corrector. The scale at the focal plane is
$38.7^{\prime\prime}$/mm ($0.57^{\prime\prime}$/pixel). On the 10-chip
CCD camera the detected field of view is $1.32^{\circ} \times
1.65^{\circ}$ = 2.18 square degrees.

\section{Telescope installation and commissioning}

The construction of the telescope was commenced in 2003. The contract
for the primary mirror went to Littkarno in Moscow. The material is
Astrosital low expansion ceramic. The mirror was figured and polished
at the Lomo company in St Petersburg. The CCD camera was designed and
constructed at the STE Laboratory of Nagoya University. It has a closed
cycle cooling unit and 10 butted EEV 2k~$\times$~4k thinned
back-illuminated CCDs. The pixel size is 15~$\mu$m square and each chip
is an EEV type CCD44-82. Four corrector lenses are required in the
light path in front of the final image plane to correct for astigmatism
and coma. All are from BK7 glass and have spherical surfaces. The
lenses were fabricated at IRL in N.Z.

Three filters are available for the microlensing observations. They
were made by Custom Scientific in Phoenix, AZ, and are 270~mm in
diameter, 15~mm thick. They correspond to Bessell V, Bessell I and MOA
broad-band red (90 per cent transmission from 632 to 860~nm). The
filters are mounted between lenses \#2 and \#3 near the prime focus.

The Nishimura company completed the telescope by mid-2004 and after a
trial assembly it was disassembled and shipped to New Zealand. A 9-m
diameter dome was also built by Nishimura. At Mt John the building was
commenced in June 2004. The dome was installed in September of that
year and the telescope in October. The telescope had a formal opening
ceremony on 1 December 2004, but the prime focus corrector lenses were
not installed until early 2005.

First light was on 2005 April 14. Since then the telescope has been
undergoing commissioning tests which have involved aligning the optical
elements, and minor adjustments to the CCD camera to improve its
performance. At the time of writing (mid-2005), routine operation of
the telescope for the MOA project is imminent.

\vspace*{-4mm}
\section{The MOA project: recent results and future goals}

Some highlights of the MOA project in the last few years, based mainly
on observations with the smaller 60-cm f/6.25 Cassegrain telescope at
Mt John, are noted here. They include:
\begin{enumerate}
\item Detection of the first ever confirmed extrasolar planet by
microlensing  (with OGLE group, 2003) (MOA-2003-BLG-53) (Bond et al.\
2004)

\item Analysis of high magnification event MOA-2003-BLG-32
  to show the absence of low-mass planets over a wide region of space
  around the lens star (Abe et al.\ 2004)

\item Measurement of the microlensing optical depth towards the galactic bulge (Sumi et
al.\ 2003)

\item Analysis of limb darkening of the source star in the microlensing
event MOA-2002-BLG-33 (Abe et al.\ 2003)

\item Analysis of stellar shape, also in MOA-2002-BLG-33 (Rattenbury et
al.\ 2005)

\item Contribution to the discovery of the second extrasolar planet
discovered by microlensing, OGLE-2005-BLG-71 (Udalski et al.\ 2005)

\item Detection of 12 transiting objects ranging in size from 1.7 -–
3.2 $R_{\rm J}$ using MOA data obtained 2000--02; they are possible new
planets (hot Jupiters), periods 1 -– 4 days (Abe et al.\ 2005)
\end{enumerate}
It is  noted that the MOA 0.6-m telescope contributed most of the data
to the first confirmed planet by microlensing (Bond et al.\ 2004) as
well as to a possible earlier event involving a lens with a low mass
planet in 1998 (Bond et al.\ 2002). The MOA 1.8-m telescope has already
made observations that are likely to contribute to new planetary events
in the near future.

In the future the MOA project will continue a vigorous campaign on
detecting and announcing new microlensing alerts. MOA will follow up
these alerts so as to find new extrasolar planets, with the assistance
of our world-wide collaborators, and hence characterize the properties
of Jovian and terrestrial mass planets orbiting other stars.

\label{lastpage}

\clearpage

\end{document}